\documentclass[prb,twocolumn,preprintnumbers,amsmath,amssymb,superscriptaddress]{revtex4}

\usepackage[dvips]{graphicx}
\usepackage{dcolumn}
\usepackage{bm}
\usepackage{longtable}
\usepackage[dvips]{color}
\usepackage{ulem}

\begin{document}


\title{
Ferroelectric atomic displacement in multiferroic tetragonal perovskite Sr$_{1/2}$Ba$_{1/2}$MnO$_3$
}

\author{D. Okuyama}%
\email[]{okudaisu@tohoku.ac.jp}
\affiliation{Institute of Multidisciplinary Research for Advanced Materials (IMRAM), Tohoku University, Katahira 2-1-1, 
Sendai 980-8577, Japan.}
\author{K. Yamauchi}%
\affiliation{ISIR-SANKEN, Osaka University, 8-1 Mihogaoka, Ibaraki, Osaka 567-0047, Japan}
\author{H. Sakai}%
\affiliation{Department of Physics, Osaka University, Toyonaka, Osaka 560-0043, Japan}
\affiliation{PRESTO, Japan Science and Technology Agency, Kawaguchi, Saitama 332-0012, Japan}
\author{Y. Taguchi}%
\affiliation{RIKEN Center for Emergent Matter Science (CEMS), Wako 351-0198, Japan}
\author{Y. Tokura}%
\affiliation{RIKEN Center for Emergent Matter Science (CEMS), Wako 351-0198, Japan}
\affiliation{Department of Applied Physics, University of Tokyo, Tokyo 113-8656, Japan}
\author{K. Sugimoto}%
\affiliation{JASRI SPring-8, Hyogo 679-5198, Japan}
\author{T. J. Sato}%
\affiliation{Institute of Multidisciplinary Research for Advanced Materials (IMRAM), Tohoku University, Katahira 2-1-1, 
Sendai 980-8577, Japan.}
\author{T. Oguchi}%
\affiliation{ISIR-SANKEN, Osaka University, 8-1 Mihogaoka, Ibaraki, Osaka 567-0047, Japan}

\date{\today}

\begin{abstract}
We investigate the crystal structure in multiferroic tetragonal perovskite Sr$_{1/2}$Ba$_{1/2}$MnO$_3$ 
with high accuracy of the order of 10$^{-3}$~\AA\ for an atomic displacement.  
The large atomic displacement for Mn ion from the centerosymmetric position, comparable with the off-centering distortion 
in the tetragonal ferroelectric BaTiO$_3$, is observed in the ferroelectric phase ($T_\mathrm{N}$ $\leq$ $T$ $\leq$ $T_\mathrm{C}$).  
In stark contrast, in the multiferroic phase ($T$ $\leq$ $T_\mathrm{N}$), the atomic displacement for Mn ion is suppressed, 
but those for O ions are enlarged.  
The atomic displacements in the polar crystal structures are also analyzed in terms of the ferroelectric modes.  
In the ferroelectric phase, the atomic displacements are decomposed into dominant positive Slater, negative Last, and small positive Axe modes.  
The suppression of Slater and Last modes, the sign change of Last mode, and the enlargement of Axe mode are found in the multiferroic phase.  
The ferroelectric distortion is well reproduced by a first-principles calculation based on  Berry phase method, 
providing an additional information on competing mechanisms to induce the polarization, electronic $p$-$d$ hybridization 
vs. magnetic exchange-striction.  
\end{abstract}

\maketitle

\section{Introduction}

Since a large nonlinear magnetoelectric effect was found in perovskite TbMnO$_3$, 
multiferroic materials have been extensively investigated~\cite{Kimura2003,Fiebig2005,Cheong2007,Tokura2014}.  
It is well known that the electric polarization of the most multiferroic materials is farc smaller than those of conventional ferroelectric materials 
such as BaTiO$_{3}$.  
Nevertheless, rather large electric polarization among the multiferroic materials are theoretically shown in tetragonal perovskite 
BaMnO$_3$~\cite{Bhattacharjee2009,Rondinelli2009,Lee2010}.  
In the paramagnetic phase, it is prospected that the ferroelectricity is induced by the off-centering distortion of the Mn and O ions.  
Since the magnetic Mn$^{4+}$ ion directly contributes to the emergence of the ferroelectricity, the large magnetoelectric effect is expected.  
However, in fact, it was known that the hexagonal structural phase is stable in BaMnO$_3$.  
Sakai \textit{et al.} found that Sr$_{1/2}$Ba$_{1/2}$MnO$_3$ with the smaller tolerance factor is crystallized 
in the tetragonal perovskite structure and shows ferroelectricity~\cite{Sakai2011}.  
Below the magnetic phase transition temperature, the change of the crystal lattice with large reduction 
of the electric polarization is observed in tetragonal perovskite Sr$_{1/2}$Ba$_{1/2}$MnO$_3$~\cite{Sakai2011}.  

\begin{figure}
\includegraphics*[width=85mm,clip]{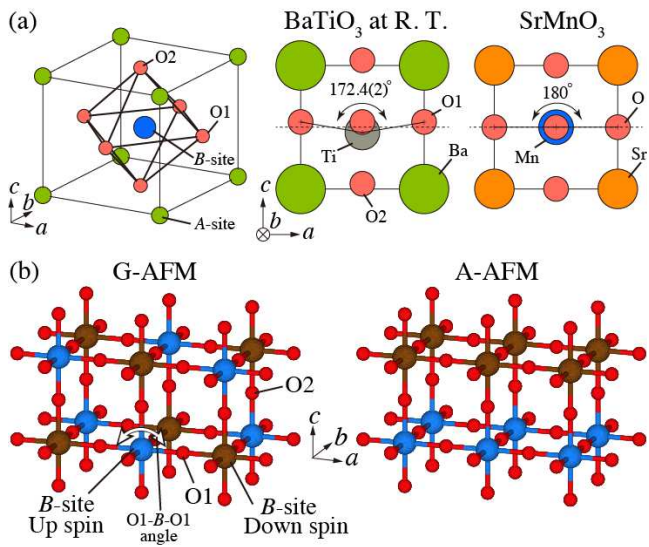}
\caption{\label{fig_01} (Color online) 
(a) Schematic illustrations of the perovskite \textit{AB}O$_3$ structure.  
The O1-\textit{B}-O1 bond angles of the BaTiO$_3$ and SrMnO$_3$ are also shown.  
(b) Schematics to explain the G-type antiferromagnetic (G-AFM) and A-type antiferromagnetic (A-AFM) structures on \textit{AB}O$_3$, 
drawn by VESTA\cite{VESTA}.  
Here, the \textit{A}-site ions are not shown.  
Blue and brown circles stand for the up and down spins on \textit{B}-site, respectively.  
O1 and O2 sites are indicated by the red circles.  
In G-AFM, the neighbor magnetic moments are aligned antiparallel.  
In A-AFM, the magnetic moments are aligned parallel in $a$-$b$ plane with antiparallel coupling with neighbor planes.  
}
\end{figure}

Perovskite Sr$_{1/2}$Ba$_{1/2}$MnO$_3$ exhibits two phase transitions, ferroelectric and antiferromagnetic, at $T_{\mathrm{C}}$ $\sim$ 400~K 
and $T_{\mathrm{N}}$ $\sim$ 185~K~\cite{Sakai2011}, respectively.  
Here, we call the phases for $T_{\mathrm{N}}$ $\leq$ $T$ $\leq$ $T_{\mathrm{C}}$ as ferroelectric and 
for $T$ $\leq$ $T_{\mathrm{N}}$ as multiferroic, respectively.  
Below $T_{\mathrm{C}}$, it is reported that the crystal system changes from the centrosymmetric cubic to the polar tetragonal, 
determined from the temperature variation of the $c$/$a$-lattice-constant ratio.  
The $c$/$a$ increases with decreasing temperature below $T_{\mathrm{C}}$ and saturates near 250~K.  
In contrast, below $T_{\mathrm{N}}$, the $c$/$a$ turns to decrease with decreasing temperature and saturates below 150~K.  
From the reduction of $c$/$a$, it was speculated that the electric polarization is also reduced.  
The crystal structure analyses with use of the twin-free single crystal was also performed in the ferroelectric phase~\cite{Sakai2011}.  
The result indicates that the origin of the ferroelectricity is the off-centering distortion of O1-Mn-O1 bond angle, 
same with the tetragonal ferroelectric BaTiO$_3$ [See Fig.~\ref{fig_01} (a)]~\cite{Hewat1973,Frazer1955,Shirane1957}.  
In the multiferroic phase, the distortion of O1-Mn-O1 at 50~K is smaller than that at 225~K in the ferroelectric phase.  
In view of the phase continuity from the antiferromagnetic ordered phase of SrMnO$_3$~\cite{Takeda1974,Chmaissem2001}, 
the antifferomagnetic structure in the multiferroic phase is inferred to be G-type (G-AFM), 
in which nearest neighbor magnetic moments are aligned antiparallel as shown in Fig.~\ref{fig_01} (b).  
In the earlier study, it was speculated that the ferroelectric polarization is suppressed to obtain the gain of the magnetic exchange 
energy~\cite{Sakai2011,Giovannetti2012}.  
To our best knowledge, however, the quantitative comparison between the experimental result and the theoretical calculation 
for the electric polarization in the multiferroic phase has not yet been done.  
To quantitatively discuss the suppression mechanism of the ferroelectric polarization, 
information of the atomic displacements and the frozen ferroelectric modes in the ferroelectric and multiferroic phases is necessary.  
Nonetheless, in the earlier crystal structure analysis, the obtained atomic displacements of the ions in multiferroic phase 
is smaller than the experimental uncertainties~\cite{Sakai2011}.  
Thus, the accurate crystal structure analysis in the multiferroic phase of tetragonal perovskite Sr$_{1/2}$Ba$_{1/2}$MnO$_3$ has been desired.  

In a multiferroic system, an essential contribution of the quantum Berry phase of valence electrons can be revealed by 
the combined study of accurate structural analysis and first-principles calculation~\cite{Okuyama2011}.  
For tetragonal perovskite Sr$_{1/2}$Ba$_{1/2}$MnO$_3$, Giovannetti \textit{et al.} performed the first-principles density functional theory 
(DFT) band simulation and claimed that the ferroelectric polarization caused by Mn-O2 hybridization is suppressed by Mn-O1-Mn superexchange 
interaction in the G-AFM ordering~\cite{Giovannetti2012}.  
In their study, the crystal structure in the multiferroics phase was theoretically optimized with generalized gradient approximation (GGA) potential 
whereas the comparison with the experimental structure was missing.  
It is also noteworthy here that a simulation study can provide an ideal magnetic structure that enhances ferroelectricity.  
In the multiferroic materials, the change of the magnetic structure may induce 
much larger ferroelectric polarization~\cite{Prokhnenko2007,Ishiwata2010}.  
Thus, for the further understanding of the multiferroic properties in tetragonal perovskite Sr$_{1/2}$Ba$_{1/2}$MnO$_3$, 
it is important as well to evaluate the ferroelectric polarization in hypothetical magnetic structures.  

In this paper, we report the atomic displacements in the ferroelectric and multiferroic phases of tetragonal 
perovskite Sr$_{1/2}$Ba$_{1/2}$MnO$_3$ determined by the crystal structure analyses using the twin-free single crystal 
and higher-\textit{Q} diffraction data than earlier work~\cite{Sakai2011}.  
By the ferroelectric mode analyses, the polar crystal structures in the ferroelectric and multiferroic phases 
for Sr$_{1/2}$Ba$_{1/2}$MnO$_3$ and other tetragonal perovskite (\textit{AB}O$_3$) were classified.  
By a first-principle calculation based on the accurate-crystal-structural parameters, we quantitatively clarify 
the suppression mechanism of the ferroelectric polarization in multiferroic phase 
and discuss the possible magnetic structure that enhances the electric polarization.  

\section{Experimental and computational procedures}

A single crystal of tetragonal perovskite Sr$_{1/2}$Ba$_{1/2}$MnO$_3$ was synthesized by a high-pressure treatment 
on the precursor sample of oxygen-deficient single crystals~\cite{Sakai2011}.  
A synchrotron x-ray diffraction experiment was performed on BL02B1 at SPring-8, Japan~\cite{Sugimoto2010}.  
The photon energy of the incident x rays was tuned at 35.04~keV.  
Using the high energy x-ray, we can access diffraction peaks with high spatial resolution up to $Q \sim$ 30~\AA$^{-1}$.  
The single crystal was crushed into cubes with a typical dimension of about 20~$\mu$m.  
The absorption coefficient $\mu$ is calculated to be 37.85~cm$^{-1}$.  
The empirical absorption correction was carried out~\cite{Higashi1995}.  
Rapid-Auto program (Rigaku Corp.) was used to obtain an F-table.  
CRYSTAL STRUCTURE (Rigaku Corp.) program was used for analyzing the crystal structure from the F-table.  
In the crystal structure analysis in the multiferroic phase at 50~K, the isotropic atomic displacement parameter $B_{\mathrm{iso}}$ 
was used for the Ba/Sr site.  

First-principles calculations were performed using the VASP code~\cite{Kresse1996} within the 
GGA + $U$~\cite{Anisimov1997} formalism with various $U$ values.  
In addition, we employed the Heyd-Scuseria-Ernzerhof (HSE06) screened hybrid functional method~\cite{HSE2003}, 
which mixes the exact non-local Fock exchange and the density-functional parametrized exchange.  
The HSE06 is known to improve the evaluation of the band gap energy and the structural distortion, 
with respect to GGA + $U$ approaches~\cite{HSEslv}.  
The cut-off energy for the plane-wave expansion of the wavefunctions was set to 400~eV and a $\boldsymbol{k}$-point 
shell of (4, 4, 3) was used for the Brillouin zone integration according to Monkhorst-Pack special point mesh.  
The crystal structure was optimized with respect to internal atomic coordinates until the remaining forces were less than 1~meV/\AA\ while 
the lattice parameters were kept at the experimental values.  

\begin{figure}
\includegraphics*[width=85mm,clip]{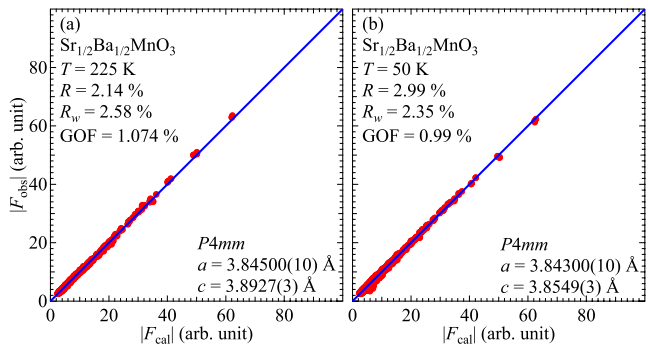}
\caption{\label{fig_02} (Color online) 
Comparison between observed ($|F_{\mathrm{obs}}|$) and calculated ($|F_{\mathrm{cal}}|$) structure factors 
(a) at 225~K in the ferroelectric and (b) at 50~K in the multiferroic phases of tetragonal perovskite Sr$_{1/2}$Ba$_{1/2}$MnO$_3$. 
}
\end{figure}

\begin{figure}
\includegraphics*[width=85mm,clip]{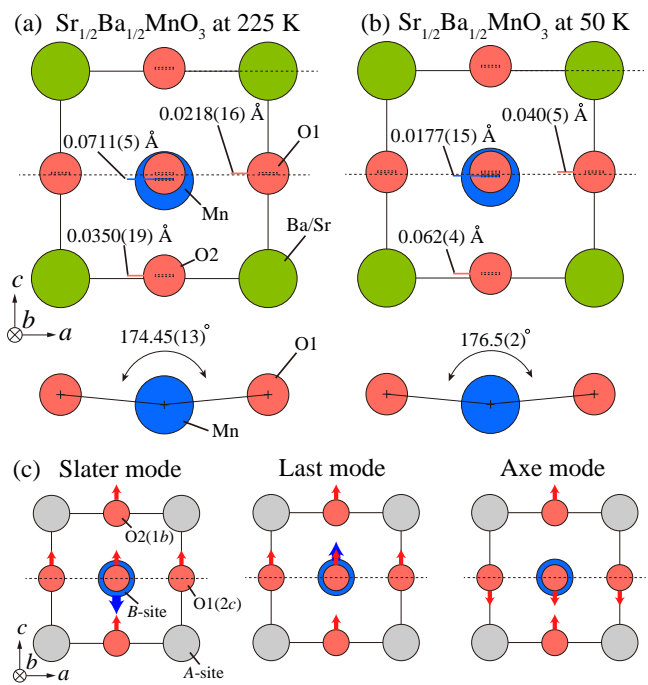}
\caption{\label{fig_03} (Color online) 
(a, b) The atomic displacements and O1-Mn-O1 bond angles of tetragonal perovskite Sr$_{1/2}$Ba$_{1/2}$MnO$_3$ at 225~K 
in the ferroelectric phase (a) and 50~K in the multiferroic phase (b).  
(c) Atomic displacements of the respective ferroelectric modes on the perovskite \textit{AB}O$_3$.  
Red and blue arrows stand for the directions of the displacement for the O and \textit{B}-site ions, respectively.  
}
\end{figure}

\section{Results and discussion}

\subsection{Synchrotron x-ray diffraction and crystal structure analysis}

The synchrotron x-ray diffraction experiments have been carried out in the ferroelectric ($T$ = 225~K) 
and in the multiferroic ($T$ = 50~K) phases of tetragonal perovskite Sr$_{1/2}$Ba$_{1/2}$MnO$_3$.  
All observed diffraction spots can be indexed by those of the $P4mm$ space group.  
By using these data sets, we performed crystal structure analyses.  
Here, the \textit{A}-site ion is fixed at the centrosymmetric position.  
The comparisons between observed and calculated structure factors are shown in Fig.~\ref{fig_02}.  
The structural parameters at 225~K and 50~K are summarized in Tables~\ref{table_1} and \ref{table_2}, respectively.  
Schematic views of the atomic displacements in the ferroelectric and multiferroic phases of tetragonal perovskite Sr$_{1/2}$Ba$_{1/2}$MnO$_3$ 
are shown in Figs.~\ref{fig_03}(a) and \ref{fig_03}(b).  
As a reference, the O1-Mn-O1 bond angles of tetragonal BaTiO$_3$ and cubic SrMnO$_3$ are also shown in Fig.~\ref{fig_01}(a).  
In the ferroelectric phase of tetragonal perovskite Sr$_{1/2}$Ba$_{1/2}$MnO$_3$, the atomic displacements along the $c$-axis at 225~K 
are respectively 0.0711(5)~\AA\ for Mn ion, 0.0218(16)~\AA\ for O1 (2$c$ site), and 0.0350(19)~\AA\ for O2 (1$b$ site), 
which are of the same order of magnitude as those reported by the earlier study~\cite{Sakai2011}.  
The off-centering distortion can be estimated by the O1-Mn-O1 bond angle, as 174.45(13)$^{\circ}$ at 225~K in the ferroelectric phase, 
comparable with that of tetragonal BaTiO$_3$~\cite{Kwei1993}.  

\begin{table*}
\caption{\label{table_1}
Structure parameters of Sr$_{1/2}$Ba$_{1/2}$MnO$_3$ at 225~K in the ferroelectric phase (Space group \textit{P}4\textit{mm} (No. 99)).  
The 22953 reflections were observed, and 3455 of them are independent.  
The 17 variables were used for the refinement.  
The lattice parameters are $a$ = 3.84500(10)~\AA\ and $c$ = 3.8927(3)~\AA.  
The reliability factors are $R$ =  2.14\%, $R_{w}$ = 2.58\%, GOF(Goodness of fit) = 1.074.  
In the tables, $x$, $y$, and $z$ are the fractional coordinates.  
Anisotropic atomic displacement parameters are represented as $U_{11}$, $U_{22}$, $U_{33}$, $U_{12}$, $U_{13}$, and $U_{23}$.  
}
\begin{ruledtabular}
\begin{tabular}{ccccccccc}
 &site&$x$&$y$&$z$&$B_{\mathrm{eq}}$&occupancy\\
\hline
Ba/Sr & 1$a$ & 0 & 0 & 0 & 0.36(5) & 1/2\\
Mn & 1$b$ & 1/2 & 1/2 & 0.48173(13) & 0.3032(15) & 1\\
O1 & 2$c$ & 0& 1/2 & 0.5056(4) & 0.609(7) & 1\\
O2 & 1$b$ & 1/2 & 1/2 & 0.0090(5) & 0.658(8) & 1\\
\end{tabular}
\begin{tabular}{ccccccc}
 &$U_{11}$ (\AA$^2$)&$U_{22}$ (\AA$^2$)&$U_{33}$ (\AA$^2$)&$U_{12}$ (\AA$^2$)&$U_{13}$ (\AA$^2$)&
 $U_{23}$ (\AA$^2$)\\
\hline
Ba & 0.00529(7) & 0.00529(7) & 0.00313(11) & 0 & 0 & 0\\
Sr & 0.00409(9) & 0.00409(9) & 0.00009(8) & 0 & 0 & 0\\
Mn & 0.00450(3) & 0.00450(3) & 0.00252(6) & 0 & 0 & 0\\
O1 & 0.00465(13) & 0.00908(19) & 0.0094(2) & 0 & 0 & 0\\
O2 & 0.00935(19) & 0.00935(19) & 0.0063(2) & 0 & 0 & 0\\
\end{tabular}
\end{ruledtabular}
\end{table*}

\begin{table*}
\caption{\label{table_2}
Structure parameters of Sr$_{1/2}$Ba$_{1/2}$MnO$_3$ at 50~K in the multiferroic phase (Space group \textit{P}4\textit{mm} (No. 99)).  
The 23321 reflections were observed, and 3409 of them are independent.  
The 14 variables were used for the refinement.  
The lattice parameters are $a$ = 3.84300(10)~\AA\ and $c$ = 3.8549(3)~\AA.  
The reliability factors are $R$ = 2.99\%, $R_{w}$ = 2.35\%, GOF(Goodness of fit) = 0.99.  
In Ba/Sr site, the isotropic atomic displacement parameter is used for the crystal structure analysis.  
}
\begin{ruledtabular}
\begin{tabular}{ccccccc}
 &site&$x$&$y$&$z$&$B_{\mathrm{iso}}/B_{\mathrm{eq}}$&occupancy\\
\hline
Ba/Sr & 1$a$ & 0 & 0 & 0 & 0.31(5) & 1/2\\
Mn & 1$b$ & 1/2 & 1/2 & 0.4954(4) & 0.298(2) & 1\\
O1 & 2$c$ & 0& 1/2 & 0.5105(13) & 0.500(8) & 1\\
O2 & 1$b$ & 1/2 & 1/2 & 0.0161(11) & 0.472(12) & 1\\
\end{tabular}
\begin{tabular}{ccccccc}
 &$U_{11}$ (\AA$^2$)&$U_{22}$ (\AA$^2$)&$U_{33}$ (\AA$^2$)&$U_{12}$ (\AA$^2$)&$U_{13}$ (\AA$^2$)&
 $U_{23}$ (\AA$^2$)\\
\hline
Mn & 0.00355(3) & 0.00355(3) & 0.00424(15) & 0 & 0 & 0\\
O1 & 0.00414(16) & 0.0068(2) & 0.0080(3) & 0 & 0 & 0\\
O2 & 0.00680(19) & 0.00680(19) & 0.0043(5) & 0 & 0 & 0\\
\end{tabular}
\end{ruledtabular}
\end{table*}

\begin{table*}
\caption{\label{table_03}
Frozen ferroelectric modes estimated from the atomic displacements for \textit{B}-site and O ions in the tetragonal ferroelectric 
and multiferroic phases.  
$S_{\mathrm{Slater}}$, $S_{\mathrm{Last}}$, and $S_{\mathrm{Axe}}$ stand for the coefficients of Slater, Last, and Axe ferroelectric modes, 
respectively.  
The contribution ratio from the $|S_{\mathrm{Slater}}|$, $|S_{\mathrm{Last}}|$, and $|S_{\mathrm{Axe}}|$ are also shown in brackets.  
$\xi_{B}$, $\xi_{\mathrm{O1}}$, $\xi_{\mathrm{O2}}$ are respectively the atomic displacements for \textit{B}, O1, and O2 sites.  
Here, we selected the sign of the atomic displacement $\xi_{B}$ so that the $S_{\mathrm{Slater}}$ is positive.  
$\angle_{\mathrm{O1}B\mathrm{O1}}$ stands for the distortion of O1-\textit{B}-O1 bond angle.  
}
\begin{ruledtabular}
\begin{tabular}{ccccccccc}
 & $\xi_{B}$ (\AA) & $\xi_{\mathrm{O1}}$ (\AA) & $\xi_{\mathrm{O2}}$ (\AA) & $S_{\mathrm{Slater}}$ & $S_{\mathrm{Last}}$ 
 & $S_{\mathrm{Axe}}$ & $\angle_{\mathrm{O1}B\mathrm{O1}}$ (deg)\\
\hline
Sr$_{1/2}$Ba$_{1/2}$MnO$_3$ & -0.0711(5) & 0.0218(16) & 0.0350(19) & 0.0519(13) & -0.0134(9) & 0.009(2) & 174.45(13) \\
(225 K) & & & & [70 \%] & [18 \%] & [12 \%] & \\
\hline
Sr$_{1/2}$Ba$_{1/2}$MnO$_3$ & -0.0177(15) & 0.040(5) & 0.062(4) & 0.035(4) & 0.007(2) & 0.014(5) & 176.5(2) \\
(50 K) & & & & [63 \%] & [13 \%] & [25 \%] & \\
\hline
BaTiO$_3$ [\onlinecite{Kwei1993}] & -0.091(2) & 0.042(2) & 0.0985(16) & 0.076(2) & -0.0087(17) & 0.037(2) & 172.4(2) \\
(300 K) & & & & [62 \%] & [7 \%] & [30 \%] & \\
\hline
KNbO$_3$ [\onlinecite{Hewat1973}] & -0.09(4) & 0.08(4) & 0.07(4) & 0.11(5) & -0.008(19) & -0.01(5) & 170.2(3) \\
(543 K) & & & & [86 \%] & [6 \%] & [8 \%] & \\
\hline
PbTiO$_3$ [\onlinecite{Nelmes1985}] & 0.1567(16) & 0.4879(13) & 0.4646(13) & 0.1615(15) & 0.2178(12) & -0.0155(15) & 160.7(2) \\
(295 K) & & & & [41 \%] & [55 \%] & [4 \%] & \\
\hline
BiCoO$_3$ [\onlinecite{Belik2006}] & 0.316(4) & 1.086(2) & 0.961(2) & 0.402(4) & 0.425(3) & -0.084(3) & 135.1(2) \\
(300 K) & & & & [44 \%] & [47 \%] & [9 \%] & \\
\end{tabular}
\end{ruledtabular}
\end{table*}

The atomic displacements and O1-Mn-O1 distortion at 50~K in the multiferroic phase are respectively changed 
as 0.0177(15)~\AA\ for Mn ion, 0.040(5)~\AA\ for O1, 0.062(4)~\AA\ for O2, and 176.5(2)$^{\circ}$.  
In this study, since the atomic displacements are determined with the accuracy of the 10$^{-3}$~\AA\ order, 
we observed that the atomic displacements of O ions are larger than that of Mn ion in the multiferroic phase.  
This enlarged atomic displacements of the O ions can not be explained only by the suppression of the off-centering distortion.  
The reason for this enlargement will be discussed later.  

Next, we analyzed the observed atomic displacements by the ferroelectric modes 
to compare the ferroelectric and multiferroic phases of tetragonal perovskite Sr$_{1/2}$Ba$_{1/2}$MnO$_3$ 
with other ferroelectric perovskite materials.  
At the structural phase transition from cubic $Pm3m$ to tetragonal $P4mm$ in the perovskite oxide, the polar vibrational motion 
is decomposed by three modes, so-called Slater, Last, and Axe modes [See Fig.~\ref{fig_03}(c)].  
The analysis by these ferroelectric modes is commonly performed to classify the soft phonon mode obtained from the optical, x-ray, 
and neutron spectroscopy experiments~\cite{Cochran1960,Shirane1970,Nunes1971,Harada1971,Scott1974,Sakai2012}.  
In this paper, we used polar atomic displacements from centrosymmetric positions to estimate frozen ferroelectric modes.  
The ratio of the frozen ferroelectric modes in tetragonal perovskite Sr$_{1/2}$Ba$_{1/2}$MnO$_3$ are compared with those 
in other ferroelectrics in Table~\ref{table_03}.  
The frozen ferroelectric modes can be quantified from the masses 
and the atomic displacements from the centrosymmetric positions of ions, 
as Harada \textit{et al.} did using the inelastic structure factor of the soft phonon modes~\cite{Harada1970}.  
Here, the polar atomic displacements of \textit{B}-site, O1, and O2 sites are represented 
by $\xi_{\mathit{B}}$, $\xi_{\mathrm{O1}}$, and $\xi_{\mathrm{O2}}$, respectively.  
The coefficients of the ferroelectric modes, $S_{\mathrm{Slater}}$, $S_{\mathrm{Last}}$, and $S_{\mathrm{Axe}}$, can be defined as 
\begin{equation}
\xi_{x} = S_{\mathrm{Slater}}\cdot\bm{s}_{\mathrm{Slater}} + S_{\mathrm{Last}}\cdot\bm{s}_{\mathrm{Last}} 
+ S_{\mathrm{Axe}}\cdot\bm{s}_{\mathrm{Axe}}.  
\end{equation}
\begin{equation}
\xi_{x} = ( \xi_{\mathit{B}}, \ \xi_{\mathrm{O1}}, \ \xi_{\mathrm{O2}} ).  
\nonumber
\end{equation}
Here, $\bm{s}_{\mathrm{Slater}} = (-k, 1, 1)$, $\bm{s}_{\mathrm{Last}} = (1+k', 1+k', 1+k')$, $\bm{s}_{\mathrm{Axe}} = (0, -1/2, 1)$, 
$k = 3M_{\mathrm{O}}/M_{\mathit{B}}$, and $k' = (M_{\mathit{B}}+3M_{\mathrm{O}})/M_{\mathit{A}}$.  
$M_{\mathit{A}}$, $M_{\mathit{B}}$, and $M_{\mathrm{O}}$ stand for the mass of \textit{A}-site, \textit{B}-site, and O ions, respectively.  
Thus, 
\begin{equation}
\left(\begin{array}{@{\,}c@{\,}}
        \xi_{\mathit{B}} \\
        \xi_{\mathrm{O1}} \\
        \xi_{\mathrm{O2}} \\
       \end{array} \right)=
\left(
       \begin{array}{@{\,}ccc@{\,}}
        - k & 1 + k' & 0 \\
        1 & 1 + k' & - \frac{1}{2} \\
       1 & 1 + k' & 1 \\
       \end{array}
       \right) 
\left(\begin{array}{@{\,}c@{\,}}
        S_{\mathrm{Slater}} \\
        S_{\mathrm{Last}} \\
        S_{\mathrm{Axe}} 
       \end{array} \right),  
\end{equation}
\begin{equation}
\left(\begin{array}{@{\,}c@{\,}}
        S_{\mathrm{Slater}} \\
        S_{\mathrm{Last}} \\
        S_{\mathrm{Axe}} 
       \end{array} \right) = 
\alpha \left(\begin{array}{@{\,}c@{\,}}
        \xi_{\mathit{B}} \\
        \xi_{\mathrm{O1}} \\
        \xi_{\mathrm{O2}} \\
       \end{array} \right), 
       \mathrm{and }
\end{equation}
\begin{equation}
\alpha = 
\frac{1}{3(1+k)}
\left(
       \begin{array}{@{\,}ccc@{\,}}
        - 3 & 2 & 1 \\
        \frac{3}{1+k'} & \frac{2k}{1+k'} & \frac{k}{1+k'} \\
       0 & -2(1+k) & 2(1+k) \\
       \end{array}
       \right).  
\nonumber
\end{equation}
In Table~\ref{table_03}, the coefficients of the ferroelectric modes are summarized for tetragonal perovskite Sr$_{1/2}$Ba$_{1/2}$MnO$_3$, 
in comparison for other perovskite ferroelectrics, tetragonal BaTiO$_{3}$, KNbO$_{3}$, PbTiO$_{3}$, 
and BiCoO$_{3}$~\cite{Hewat1973,Kwei1993,Nelmes1985,Belik2006}.  

In the ferroelectric phase of tetragonal perovskite Sr$_{1/2}$Ba$_{1/2}$MnO$_3$, the dominant positive 
$S_{\mathrm{Slater}}$, relatively large negative $S_{\mathrm{Last}}$, and small positive $S_{\mathrm{Axe}}$ are obtained.  
The contributions from the $|S_{\mathrm{Slater}}|$, $|S_{\mathrm{Last}}|$, and $|S_{\mathrm{Axe}}|$ are approximately 
70~\%, 18~\%, and 12~\%, compatible with the result (71~\%, 24~\%, and 5~\%) obtained by the optical 
and inelastic x-ray spectroscopies~\cite{Sakai2012}.  
In tetragonal BaTiO$_3$, the dominant positive $S_{\mathrm{Slater}}$, small negative $S_{\mathrm{Last}}$, 
and relatively large positive $S_{\mathrm{Axe}}$ are observed.  
In tetragonal KNbO$_3$, the positive $S_{\mathrm{Slater}}$ is dominant but the error bars for the other modes are too large.  
As a commonality, they share two characteristics, the dominant positive $S_{\mathrm{Slater}}$ and negative $S_{\mathrm{Last}}$.  

To explain the origin of the commonality, we refer the earlier first-principles calculations for perovskite oxides, 
which have pointed out the importance of the covalency between the \textit{B}-site and the apical O2 ions for the emergence of 
ferroelectricity~\cite{Cohen1992_1,Cohen1992_2,Miyazawa2000,Chasse2011}.  
This is the reason why the dominant parameter is the Slater mode with contracting the distance between 
the \textit{B}-site and O2 ions as shown in Fig.~\ref{fig_03}(c).  
The negative $S_{\mathrm{Last}}$ and positive $S_{\mathrm{Axe}}$ play a role in reducing the extra atomic displacements of the O1 ions 
generated by the Slater mode.  

In PbTiO$_3$, the earlier first-principles calculation also pointed out that the hybridization between the 6$p$ band of Pb and 2$p$ band of O1 
induces additional component of the electric polarization~\cite{Miyazawa2000}.  
In that case, the distance between the Pb and O1 ions also decreases.  
This atomic displacement induces the combined $S_{\mathrm{Slater}}$ and $S_{\mathrm{Last}}$ mode, 
which can be actually seen in PbTiO$_3$, as listed in Table~\ref{table_03}.  
In BiCoO$_3$, since the ratio of the ferroelectric modes is similar with that of PbTiO$_3$, 
we speculate that the origin of the ferroelectricity for BiCoO$_3$ is the same for PbTiO$_3$.  

In the multiferroic phase of Sr$_{1/2}$Ba$_{1/2}$MnO$_3$, $S_{\mathrm{Slater}}$ and $S_{\mathrm{Last}}$ are suppressed, 
while $S_{\mathrm{Axe}}$ is enlarged.  
In addition, the sign of $S_{\mathrm{Last}}$ changes to positive.  
The G-AFM exchange interaction prefers 180$^{\circ}$ O1-Mn-O1 bond angle, 
being contradictory with off-centering distortion of O1-Mn-O1 bond angle.  
Therefore, the displacement for Mn ion is suppressed, and consequently gives rise to the decrease of $S_{\mathrm{Slater}}$ 
and $S_{\mathrm{Last}}$ modes.  
In stark contrast, the apical O2 is relatively free from the restriction of the magnetic exchange interaction.  
Thus, we speculate that the atomic displacement for apical O2 is enlarged to obtain the gain of the covalency between Mn and O2, 
resulting in the enlarged $S_{\mathrm{Axe}}$ parameter.  
To eliminate the extra atomic displacements of the O1 from $S_{\mathrm{Axe}}$, 
the sign of $S_{\mathrm{Last}}$ mode changes to positive in the multiferroic phase.  
From the experimentally determined crystal structure information and the results of the mode analyses, 
we discussed and speculated the qualitative suppression mechanism of the ferroelectricity.  
To support this speculation and provide more quantitative discussion, we performed the first-principles calculation.  

\subsection{First-principles calculation}

For the discussion of the atomic displacements and the resulting ferroelectricity in this system, 
we performed first-principles calculations for tetragonal perovskite Sr$_{1/2}$Ba$_{1/2}$MnO$_3$.  
To understand the effect of the magnetic order upon the ferroelectricity in the multiferroic phase, 
here we also simulate the ferroelectric polarization in the hypothetical A-type antiferromagnetic (A-AFM) structure 
(the magnetic moments are aligned parallel in $a$-$b$ plane with antiparallel coupling with neighbor planes as shown 
in Fig.~\ref{fig_01}(b)) as well as the ground-state G-AFM structure in Sr$_{1/2}$Ba$_{1/2}$MnO$_3$.  

\begin{figure}
\includegraphics*[width=80mm,clip]{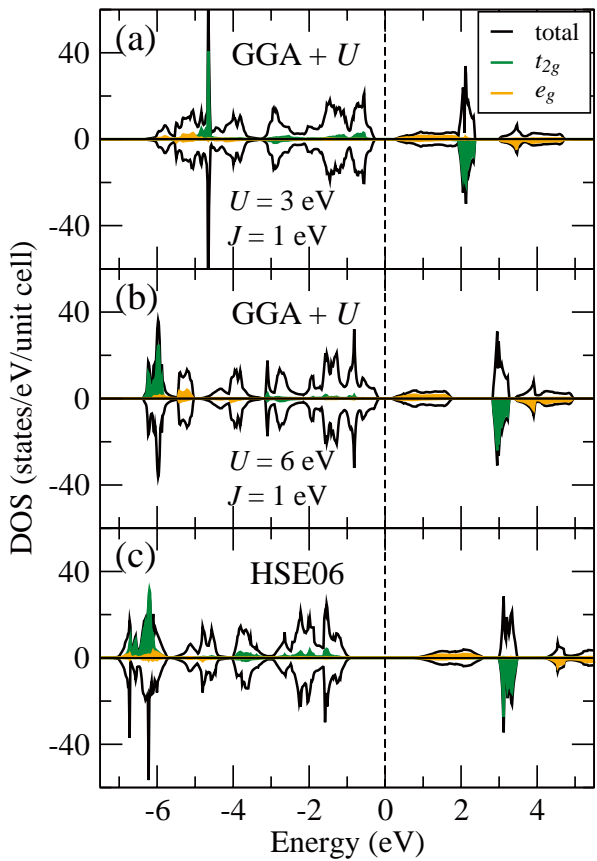}
\caption{\label{fig_04} 
Density of states (DOS) in G-AFM of Sr$_{1/2}$Ba$_{1/2}$MnO$_3$ calculated from GGA + $U$ method with (a) $U$ = 3~eV and $J$ = 1~eV, 
(b) $U$ = 6~eV and $J$ = 1~eV, and from (c) HSE06 method.  
Upper and lower panels show majority- and minority-spin states, respectively.  
Projected DOS for Mn $t_{2g}$ and $e_{g}$ orbital states are highlighted with green and orange colors, respectively.  
}
\end{figure}

Figures~\ref{fig_04}(a) and \ref{fig_04}(b) show the density of states from GGA + $U$ calculations.  
When we set $U$ = 3~eV and $J$ = 1~eV as consistent with the previous DFT study~\cite{Giovannetti2012}, 
the system is insulator while the energy gap is significantly underestimated as $E_{\mathrm{gap}}\sim$ 0.5~eV, 
being inconsistent with the experimentally estimated energy gap $\sim$ 2~eV for SrMnO$_3$~\cite{Saitoh1995}.  
The underestimation of the energy gap was not improved 
when  the $U$ value was increased up to 6~eV [see Fig. \ref{fig_04}(b)]; 
on the contrary, the band gap was reduced to be $\sim$0.3 eV. 
This result might seem counterintuitive but 
this is due to a property of GGA + $U$ method that adds effective Coulomb potential only to the localized orbital states (such as 3$d$ and 4$f$ orbital states).  
Indeed, GGA + $U$ Coulomb potential shifts down the occupied Mn 3$d$ states but keeps delocalized O 2$p$ states at the original energy levels around the valence top state.  
When the O 2$p$ states are located at shallow energy level, Mn ion favors to show trivalent instead of quadrivalent ionic state.  
This is the reason why the band gap tends to be closed as increasing the $U$ value.  
To make matters worse, this narrow energy gap is closed when hypothetical ferromagnetic phase 
or A-AFM phase is calculated.  
Therefore, we conclude that GGA + $U$ approach is not appropriate to describe the wide gapped insulating state 
and evaluate the ferroelectric distortion in Sr$_{1/2}$Ba$_{1/2}$MnO$_3$.  

Figure~\ref{fig_04}(c) shows the density of states from HSE06 calculation, leading to the wider energy gap ($E_{\mathrm{gap}}\sim$ 2~eV, 
consistent with the experimental data in SrMnO$_3$~\cite{Saitoh1995}) with Mn quadrivalent state.  
In this case, the fraction of exact Hartree-Fock exchange in HSE06 scheme shifts down both the occupied Mn $3d$ levels and O $2p$ levels.  
Hereinafter, we will focus on HSE06 results and discuss the ferroelectric property.  
By using the experimental and DFT-optimized crystal structures, the ferroelectric polarization was calculated 
as listed in Table~\ref{table_04}.  
In order to investigate the influence of the magnetic ordering to the ferroelectric polarization, 
we consider the ground-state G-AFM and the hypothetical A-AFM configurations.  

It is noted that the calculated electric polarization by the optimized structure based on point-charge model 
with nominal ionic charges (Ba and Sr; 2+, Mn; 4+, O; 2-), 
{\it i.e.,} the ionic displacement contribution to the electric polarization, shows good agreement with that estimated by the experimental crystal structure: 
$P_{\mathrm{PCM}}$ $\sim$ 10.1~$\mu$C/cm$^{2}$ with G-AFM in both experimental and optimized structures at $T$ = 50~K.  
This result supports the advantage of use of HSE06 functional for the polar structural distortion with a high accuracy.  
The total electric polarization $P_{\rm Berry}$, $i.e.,$ the summation of ionic and electronic contributions, is almost double of the $P_{\mathrm{PCM}}$, as often seen in other ferroelectric manganites~\cite{Picozzi2007}, and is of the same order of magnitude 
with the electric polarization (13.5 $\mu$C/cm$^{2}$) experimentally obtained in the earlier study~\cite{Sakai2011}.  

\begin{table}
\caption{\label{table_04}
Calculated ferroelectric polarization for the experimental (E) and optimized (O) crystal structure at $T$ = 50~K 
for the G-AFM and A-AFM antiferromagnetic ordering as based on HSE06-exchange-correlation functional.  
Both the net polarization obtained by Berry phase method ($P_{\mathrm{Berry}}$) and the ionic contribution based on point charge model 
($P_{\mathrm{PCM}}$) are shown in unit of $\mu$C/cm$^{2}$.  
}
\begin{ruledtabular}
\begin{tabular}{ccccc}
 &E$_{\mathrm{G-AFM}}$&E$_{\mathrm{A-AFM}}$&O$_{\mathrm{G-AFM}}$&O$_{\mathrm{A-AFM}}$\\
\hline
$P_{\mathrm{Berry}}$ & 19.37 & 23.65 & 20.17 & 30.24\\
$P_{\mathrm{PCM}}$ & 10.05 & 10.05 & 10.11 & 15.45\\
\end{tabular}
\end{ruledtabular}
\end{table}

\begin{figure}
\includegraphics*[width=60mm,clip]{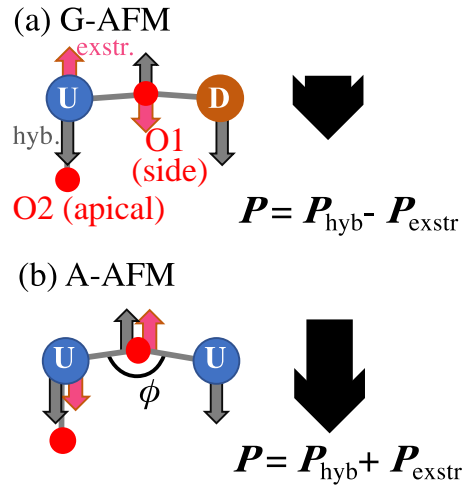}
\caption{\label{fig_05} 
Schematic illustration of ionic distortion (narrow gray and red arrows) and induced electric polarization $P$ (wide black arrows) 
for (a) G-type (G-AFM) and (b) A-type (A-AFM) antiferromagnetic orderings.  
Narrow gray and red arrow stand for the atomic displacement originating from the hybridization between Mn 3$d$ and 
apical O2 2$p$ bands and the in-plane Mn-O1-Mn magnetic magnetic exchange striction, respectively.  
$P_{\mathrm{hyb}}$ and $P_{\mathrm{exstr}}$ stand for the electric polarization from the hybridization and the exchange striction, respectively.  
U and D denote up- and down-spins Mn sites, respectively.  
The detailed crystal and magnetic structures are shown in Fig.~\ref{fig_01}(b).  
}
\end{figure}

Next, we focus on the the suppression mechanism of the ferroelectricity in the multiferroic phase.  
Basically, we consider two mechanisms to induce the polarization:  hybridization between Mn 3$d$ and apical O2 2$p$ states ($P_{\rm hyb}$), 
and in-plane Mn-O1-Mn magnetic exchange striction ($P_{\rm extr}$) as shown in Fig.~\ref{fig_05}.  
The $p-d$ hybridization drives polar ionic distortion of Slater mode, by which Mn and O2 ions are respectively shifted downward and upward.  
The magnetic exchange striction modulates the in-plane Mn-O1-Mn bond angle ($\phi$), resulting in the suppression of the $S_{\mathrm{Slater}}$ 
and the sign change of $S_{\mathrm{Last}}$.  
Since the driving mechanism of the change of the ferroelectricity upon the ferroelectric to multiferroic phase transition 
is the magnetic exchange interaction, 
the atomic displacement should depend on the Mn spin configuration.  
In G-AFM, the magnetic exchange striction favors $\phi$ = 180$^{\circ}$ so that Goodenough-Kanamori rule is satisfied 
for Mn$^{4+}$ ion~\cite{Goodenough1955,Goodenough1958,Kanamori1959}.  
This magnetic exchange striction prevents the atomic displacement of the side O1 ion so that total electric polarization is reduced.  
In contrast, in case of the hypothetical A-AFM, the magnetic exchange striction favors $\phi$ = 90$^{\circ}$ and enhances 
the hybridization-induced polarization as shown in Fig.~\ref{fig_05}.  
The calculated polarization with G-AFM and A-AFM is $P_{\mathrm{Berry}}$ = 20.17 and 30.24~$\mu$C/cm$^{2}$, respectively, 
as being consistent with the above discussed mechanism.  
The difference of the  $P$ values  allows us to decompose the polarization into two contributions, 
$P_{\rm hyb} \sim 25$ and $P_{\rm extr} \sim 5$ $\mu$C/cm$^{2}$. 
The former is comparable to the conventional ferroelectric polarization in BaTiO$_{3}$ ($P$ $\sim$ 26 $\mu$C/cm$^{2}$) 
and the latter is comparable to the magnetically-driven polarization in multiferroic HoMnO$_{3}$ 
($P$ $\sim$ 6 $\mu$C/cm$^{2}$)\cite{Picozzi2007}. 
Thus, we conclude that since only positive $P_{\rm hyb}$ contributes the ferroelectric polarization in the paramagnetic phase, 
negative $P_{\rm extr}$ causes the suppression of the polarization observed in the multiferroic phase.  
If one succeeded in stabilizing A-AFM in tetragonal \textit{A}MnO$_{3}$ system, it might be a milestone multiferroic demonstrating 
the polarization larger than representative ferroelectric BaTiO$_{3}$.  
Nonetheless, since A-AFM is energetically unfavored by 1.2~eV/f.u. with respect to G-AFM, 
a study to stabilize the A-AFM \textit{A}MnO$_{3}$ is left as a topic for future work.  

\section{Summary}

In summary, we have performed the synchrotron x-ray diffraction experiment to investigate the accurate crystal structures 
in the ferroelectric and multiferroic phases of tetragonal perovskite Sr$_{1/2}$Ba$_{1/2}$MnO$_3$ using twin-free-single-crystalline sample.  
The large atomic displacement for Mn ion was observed in the ferroelectric phase.  
In the multiferroic phase, by contrast the atomic displacement for Mn ion is suppressed, but those for O ions are enlarged.  
From the obtained crystal structural parameters, the ferroelectric mode analyses were carried out.  
In the ferroelectric phase, the atomic displacements can be decomposed as dominant positive Slater, negative Last, 
and small positive Axe modes.  
The suppression of Slater and Last modes, the sign change of Last mode, and the enlargement of Axe mode are found in the multiferroic phase.  
The first-principles calculation using HSE06 functional successfully described the wide-gap insulating electronic states and 
quantitatively reproduces the experimentally observed ferroelectric polarization.  
The calculated ferroelectric polarization is further decomposed into two parts relevant to the hybridization and exchange striction mechanisms.  

\begin{acknowledgments}
The authors are grateful to T. Arima and H. Katsumoto for fruitful discussions.  
The synchrotron x-ray diffraction experiment was performed at SPring-8 with approval of the JASRI 
(Proposal Numbers 2009B1304 and 2010A1795). 
This work was in part supported by Grant-in-Aids for Scientific Research (Nos. 17H02916, 17K14327, 19K03709, and 19H05822) 
from the Ministry of Education, Culture, Sports, Science and Technology (MEXT), Japan, 
and by the Research Program for CORE lab of "Dynamic Alliance for Open Innovation Bridging Human, 
Environment and Materials" in "Network Joint Research Center for Materials and Devices".
The computation in this work has been partially done using the facilities of the Supercomputer Center, 
the Institute for Solid State Physics, the University of Tokyo.
\end{acknowledgments}

\end{document}